\documentclass[12pt]{article}
\newcommand{\tr}{\mbox{tr}}

\begin{document}

\begin{flushright}
hep-th/0101226

January 2001

OU-HET 378
\end{flushright}

\begin{center}

\vspace{3cm}

\textbf{\large BPS Bound States of}

\textbf{\large D6-branes and Lower Dimensional D-branes }

\vspace{2cm}

Matsuo Sato \footnote{e-mail address : machan@funpth.phys.sci.osaka-u.ac.jp}

\vspace{1cm}

\textit{Department of Physics, Graduate School of Science, Osaka University, }

\textit{Toyonaka, Osaka, 560-0043, Japan}

\vspace{1cm}

\textbf{Abstract}

\end{center}
We construct 1/8, 1/4, and 1/2 BPS solutions spanned by diagonal elements of U(N) constant fluxes in the 6+1 dimensional U(N) super Yang-Mills theory on $T^6$ with topological stability. These solutions represent BPS bound states of D0, D2, D4, and D6 branes. The consistency with the D-brane charge conservation implies that unstable D0-D2-D6-brane systems in a B field decay to the BPS solutions, in which lower dimensional D-branes are dissolved in  D6-branes.        

\newpage

\section{Introduction}
Recently the fate of unstable D-brane systems has been well understood. D-brane anti-D-brane systems in type II superstring theory, non-BPS D-branes in type II, and bosonic D-branes decay to lower dimensional D-branes or the vacuum. Their behavior is analyzed by the potential of a tachyon field in low energy effective actions or in string field theories. The process is explained by the tachyon condensation. The instability is caused by a tachyonic mode of strings which end on unstable branes or one unstable brane.

 It is natural to think that if there are branes of various dimensions in a B field and tachyonic modes of strings ending on these branes, the system is unstable and decays to a stable system, a BPS bound state. The stability of Dp-Dp' system with B field is discussed in ~\cite{p-p'}. That conjecture is partially verified in a very small region of moduli of a B field : for a D0-D2-brane system in a large B field by using 1/B expansion in string amplitudes ~\cite{narain}, for a D0-D2-brane system in a large B field by using 1/B expansion in boundary string field theory ~\cite{david}, and for a D0-D6-brane system in a B field in the region near the BPS loci by using a low energy effective action on a D0-brane ~\cite{park, witten}. 
 
 We do not have a formalism which describes the fate of unstable systems composed of branes of various dimensions in a generic B field. We will search for a clue for understanding the mechanism. We consider an unstable system which consists of N coincident D6-branes in a B field and lower dimensional D-branes localized on them. If we turn on a B field on D6-branes, there appear dissolved lower dimensional D-branes with charges $ \int_2 B $ (D4-branes), $\int_4 B \wedge B $ (D2-branes), or $ \int_6 B \wedge B \wedge B $ (D0-branes) as the Chern-Simons-like term of low energy effective action has the form, $i \mu_p \int_{p+1} \tr(\exp(2 \pi \alpha'F_2 + B_2) \wedge \Sigma_q C_q)$ ~\cite{instanton, within, jabbari0-2}. Consequently N D6-branes will acquire D4-brane charges, D2-brane charges, and D0-brane charges. Dp-D(p+2)-brane bound state is D(p+2)-brane with a constant flux, in which a Dp-brane delocalizes and dissolves in a D(p+2)-brane. Therefore, the final state will be a BPS bound state, made of D6-branes with dissolved lower dimensional branes. The D0-Dp bound states are discussed recently in ~\cite{p-cycle, ohta}.
 
 We will investigate these bound states by using torons ~\cite{thooft}, which are topologically stable constant flux solutions with a twisted boundary condition of p+1 dimensional U(N) super Yang-Mills theory ~\cite{SYM} on $T^p$ ~\cite{taylor, toron}. We find the condition for topological stability in section 3. N coincident D6-branes with diagonal elements of U(N) constant fluxes are T-dual to N D3-branes at angles ~\cite{angle,jabbariangle, p-p'}. We obtain the BPS conditions by using this correspondence in section 2. We calculate brane charges, energy of the bound state, and energy of the system consists of N D6-branes and of localized lower dimensional D-branes on them in a diagonal basis of U(N) in section 4. In section 5 we construct 1/8 BPS and 7/8 SUSY breaking, 1/4 BPS and 3/4 SUSY breaking, and 1/2 BPS solutions spanned by diagonal elements of U(N) fluxes. Super symmetric torons in four dimensions were used to discuss black holes in ~\cite{perry, costa}. These solutions have lower energies than the localized brane systems. The BPS states consist of N coincident D6-branes in which lower dimensional branes are dissolved. We interpret that D0-D2-D6-brane systems in a B field ,in which the D0-branes and the D2-branes are localized on the D6-branes, will decay to the BPS solutions with the aid of the D-brane charge conservation. Energies of  D0-D2-D6-brane systems in a B field is bigger than those of the BPS states. We prove that D0-D6-brane systems in a B field can not decay to the BPS solutions in the appendix.

\section{1/8, 1/4, and 1/2 BPS States of N D6 Systems with Constant Fluxes}
There are stable D8, D6, D4, D2, and D0-branes in type IIA super string theory. In the closed string decoupling limit, D-brane effective action is a Dirac-Born-Infeld action, which is Yang-Mills theory in the $ \alpha ' \to 0 $ limit. We consider N coincident D6-branes, which are described by the U(N) super Yang-Mills theory. Bound states of N coincident Dp-branes and lower dimensional branes are described by torons~\cite{thooft} in the super Yang-Mills theory on $T^p$~\cite{taylor, toron}. Torons are classically stable static constant fluxes $F$ with a twisted boundary condition. We will seek for a class of solutions, which is 1/8, 1/4, and 1/2 BPS and is spanned by diagonal elements of U(N) constant fluxes. U(1) part of this solution can be identified with the U(1) part of the field strength plus the NS-NS B field, because these two fields are connected by gauge transformation and the DBI lagrangian has the form $ \sqrt{det(1+B+2\pi\alpha'F)}$ turned on a B field. We write $\mathcal{F}=B+2\pi\alpha'F$ in the following.
We put an ansatz,
\[\mathcal{F}_{\mu\nu}=\mathcal{F}_{\mu\nu}^av^a, \]
\[v^1=diag(10 \ldots 0), v^2=diag(010\ldots0), \ldots, v^N=diag(0\ldots01),\]
\begin{equation}
\mathcal{F}_{\mu\nu}^a=\left(
\begin{array}{@{\,}ccccccc@{\,}}
 0 & 0 & 0 & 0 & 0 & 0 & 0 \\ 
 0 & 0 & \mathcal{F}_{12}^a & 0 & 0 & 0 & 0 \\
 0 & -\mathcal{F}_{12}^a & 0 & 0 & 0 & 0 & 0 \\
 0 & 0 & 0 & 0 & \mathcal{F}_{34}^a & 0 & 0 \\
 0 & 0 & 0 & -\mathcal{F}_{34}^a & 0 & 0 & 0 \\
 0 & 0 & 0 & 0 & 0 & 0 & \mathcal{F}_{56}^a \\
 0 & 0 & 0 & 0 & 0 & -\mathcal{F}_{56}^a & 0 
\end{array}
\right). \label{eq:ansatz}
\end{equation}   
We use the $ 2\pi \alpha' = 1$ convention.

Now, let us turn to the string picture. Because gauge fields of the anzats on the coincident D6-branes have a diagonal part only, a string end on only one of them. We name each D6-brane $\mbox{D}^a$-brane, where $a$ runs from 1 to N. On D-branes Neumann boundary conditions of open string is modified when we turn on a gauge field. Of course, both a field strength $F$ and a NS-NS B field act on the boundary condition in a similar way,
\[ \partial_{\sigma} X_{\mu} +  \mathcal{F}_{\mu\nu}^a i \partial_\tau X_{\nu} = 0.\]
This is a boundary condition of an open string ending on the $\mbox{D}^a$-brane.  Putting the ansatz, we obtain
\begin{eqnarray}
\partial_{\sigma} X_1 +  \mathcal{F}_{12}^a i \partial_\tau X_{2} = 0 \nonumber \\
\partial_{\sigma} X_3 +  \mathcal{F}_{34}^a i \partial_\tau X_{4} = 0 \nonumber \\
\partial_{\sigma} X_5 +  \mathcal{F}_{56}^a i \partial_\tau X_{6} = 0. \nonumber
\end{eqnarray}
If we apply T-duality in the 2, 4, and 6 directions, Dirichlet conditions turn into Neumann conditions. N coincident D6-branes turn into D3-branes at angles.
\begin{eqnarray}
\partial_{\sigma}( X'_1 +  \mathcal{F}_{12}^aX'_{2}) = 0 \nonumber \\
\partial_{\sigma}( X'_3 +  \mathcal{F}_{34}^aX'_{4}) = 0 \nonumber \\
\partial_{\sigma}( X'_5 +  \mathcal{F}_{56}^aX'_{6}) = 0. \nonumber
\end{eqnarray}
This means that the brane directions satisfy
\begin{eqnarray}
X'_2=  \mathcal{F}_{12}^aX'_1 \nonumber \\
X'_4=  \mathcal{F}_{34}^aX'_3 \nonumber \\
X'_6=  \mathcal{F}_{56}^aX'_5. \nonumber 
\end{eqnarray}
In other words, the $a$-th brane is obtained by rotating a D3-brane extending (1,3,5). Rotate counterclockwise in the (1,2) plane by $\phi_1^a = \tan^{-1}( \mathcal{F}_{12}^a) $, in the (3,4) plane by $\phi_2^a = \tan^{-1}( \mathcal{F}_{34}^a) $, in the (5,6) plane by $\phi_3^a = \tan^{-1}( \mathcal{F}_{56}^a) $.

First of all, let us consider two D3-branes at angle. Let both initially be extended in the (1,3,5)-directions, and rotate one of them by an angle $\phi_1$ in the (1,2) plane, and so on. The existence of a D-brane breaks 1/2 SUSY. Originally unbroken super charges are
\begin{equation}
Q_\alpha + \beta^\bot \tilde{Q}_\alpha, \label{eq:scharge}
\end{equation}
where $\beta^\bot$ is $\Gamma^9\Gamma\Gamma^8\Gamma\Gamma^7\Gamma\Gamma^6\Gamma\Gamma^4\Gamma\Gamma^2\Gamma$ for a (1,3,5)-brane. $\bot$ denotes Dirichlet direction of a D-brane. After the rotation, unbroken super charges are given by
\[Q_\alpha + \rho^{-1} \beta^\bot \rho \tilde{Q}_\alpha, \]
\[\rho = \exp(i\Sigma_{i=1}^3 \phi_i s_i ), \qquad  s_i=-i/4[\Gamma^{2i-1},\Gamma^{2i}.] \] 
These two sets of the charges coincide with each other if
\[ (\beta^\bot)^{-1} \rho^{-1} \beta^\bot \rho = 1, \] 
which is equivalent to  
\[\rho^2 = 1.\]
Operator $s_a$ has two eigenvalues, 1/2 and $-$1/2. There are eight sets of eigenvalues $(s_1,s_2,s_3)$ for generic states. 1/8 BPS condition corresponds to surviving two sets of eigenvalues in eight sets, because 16 superchages (\ref{eq:scharge}) correspond to 1/2 and two of eight eigenvalue sets correspond to 1/4. The 1/8 BPS condition for angles and surviving sets of eigenvalues are given by

\hspace{30mm}

\begin{tabular}{l|l}

state & condition \\
\hline
$(+1/2,+1/2,+1/2)$ & $\phi_1 + \phi_2 + \phi_3 = 0 \quad (\mbox{mod 2} \pi)$ \\$(-1/2,-1/2,-1/2)$ &  \\
\hline
$(+1/2,+1/2,-1/2)$ & $\phi_1 + \phi_2 - \phi_3 = 0 \quad (\mbox{mod 2} \pi)$  \\
$(-1/2,-1/2,+1/2)$ & \\
\hline
$(+1/2,-1/2,+1/2)$ & $\phi_1 - \phi_2 + \phi_3 = 0 \quad (\mbox{mod 2} \pi)$ \\
$(-1/2,+1/2,-1/2)$ & \\
\hline
$(-1/2,+1/2,+1/2)$ & $-\phi_1 + \phi_2 + \phi_3 = 0 \quad (\mbox{mod 2} \pi)$ \\
$(+1/2,-1/2,-1/2)$ & \\
 
\end{tabular}

\hspace{30mm}

Let us go on to the case of N D3-branes at angles. We will consider only D3-branes which are T-dual to D6-branes with diagonal elements of U(N) constant fluxes. We can identify these D3-branes with $(\phi^a_1,\phi^a_2,\phi^a_3)$, where $a$ runs from 1 to N. Unbroken super charges of $D^1$-brane and those of $D^a$-brane coincide with each other if 
\[(\rho_a)^2 = 1 , \qquad \rho_a = \exp(i\Sigma_{i=1}^3 (\phi^a_i-\phi^1_i ) s_i) \qquad  ( a = 2, 3, \ldots, N ) \]
The condition for co-existence of $D^a$-brane and $D^b$-brane is satisfied automatically under the above conditions, as $ \exp(2i\Sigma_{i=1}^3 (\phi^a_i-\phi^b_i ) s_i) = (\rho_a)^2(\rho_b)^{-2}.$
In order to keep 1/8 super charges, two sets of eigenvalues need survive. 

\begin{tabular}{l|l}

state  & condition \\
\hline
$(+1/2,+1/2,+1/2)$  & $ (\phi^{(2)}_1-\phi^{(1)}_1) + (\phi^{(2)}_2-\phi^{(1)}_2) + (\phi^{(2)}_3-\phi^{(1)}_3) $ \\
$(-1/2,-1/2,-1/2)$ & $ =(\phi^{(3)}_1-\phi^{(1)}_1) + (\phi^{(3)}_2-\phi^{(1)}_2) + (\phi^{(3)}_3-\phi^{(1)}_3) $  \\
                 & $=\ldots $ \\
                 & $=(\phi^{(N)}_1-\phi^{(1)}_1) + (\phi^{(N)}_2-\phi^{(1)}_2) + (\phi^{(N)}_3-\phi^{(1)}_3) $ \\
                 & $=0 \quad (\mbox{mod 2} \pi)$ \\
\hline
$(+1/2,+1/2,-1/2)$  & $(\phi^{(2)}_1-\phi^{(1)}_1) + (\phi^{(2)}_2-\phi^{(1)}_2) - (\phi^{(2)}_3-\phi^{(1)}_3)$  \\
$(-1/2,-1/2,+1/2)$ & $=(\phi^{(3)}_1-\phi^{(1)}_1) + (\phi^{(3)}_2-\phi^{(1)}_2) - (\phi^{(3)}_3-\phi^{(1)}_3)$ \\
                 & $= \ldots$  \\
                 &  $= (\phi^{(N)}_1-\phi^{(1)}_1) + (\phi^{(N)}_2-\phi^{(1)}_2) - (\phi^{(N)}_3-\phi^{(1)}_3)$ \\
                 & $=0 \quad (\mbox{mod 2} \pi) $ \\
                 \hline                 
$(+1/2,-1/2,+1/2)$  & $(\phi^{(2)}_1-\phi^{(1)}_1) - (\phi^{(2)}_2-\phi^{(1)}_2) + (\phi^{(2)}_3-\phi^{(1)}_3)$ \\
$(-1/2,+1/2,-1/2)$ & $= (\phi^{(3)}_1-\phi^{(1)}_1) - (\phi^{(3)}_2-\phi^{(1)}_2) + (\phi^{(3)}_3-\phi^{(1)}_3)$ \\
                 & $= \ldots $ \\
                 & $= (\phi^{(N)}_1-\phi^{(1)}_1) - (\phi^{(N)}_2-\phi^{(1)}_2) + (\phi^{(N)}_3-\phi^{(1)}_3)$ \\
                 & $= 0 \quad (\mbox{mod 2} \pi) $ \\
\hline
$(-1/2,+1/2,+1/2)$ & $-(\phi^{(2)}_1-\phi^{(1)}_1) + (\phi^{(2)}_2-\phi^{(1)}_2) + (\phi^{(2)}_3-\phi^{(1)}_3)$  \\
$(+1/2,-1/2,-1/2)$ & $=-(\phi^{(3)}_1-\phi^{(1)}_1) + (\phi^{(3)}_2-\phi^{(1)}_2) + (\phi^{(3)}_3-\phi^{(1)}_3)$ \\
                 & $= \ldots $ \\
                 & $=-(\phi^{(N)}_1-\phi^{(1)}_1) + (\phi^{(N)}_2-\phi^{(1)}_2) + (\phi^{(N)}_3-\phi^{(1)}_3)$ \\
                 & $= 0 \quad (\mbox{mod 2} \pi)$ \\

  \end{tabular}

\hspace{30mm}
  
Finally, we obtain the 1/8 BPS condition for the configuration of N D3 branes at angles,
\begin{eqnarray}
& &\phi^{(1)}_1+\phi^{(1)}_2+\phi^{(1)}_3 = \phi^{(2)}_1+\phi^{(2)}_2+\phi^{(2)}_3 = \ldots = \phi^{(N)}_1+\phi^{(N)}_2+\phi^{(N)}_3 \quad (\mbox{mod 2} \pi) \label{eq:1/8bps1} \\
&\mbox{or}& \nonumber \\
& &\phi^{(1)}_1+\phi^{(1)}_2-\phi^{(1)}_3 = \phi^{(2)}_1+\phi^{(2)}_2-\phi^{(2)}_3 = \ldots = \phi^{(N)}_1+\phi^{(N)}_2-\phi^{(N)}_3 \quad (\mbox{mod 2} \pi) \label{eq:1/8bps2} \\
&\mbox{or}& \nonumber \\
& &\phi^{(1)}_1-\phi^{(1)}_2+\phi^{(1)}_3 = \phi^{(2)}_1-\phi^{(2)}_2+\phi^{(2)}_3 = \ldots = \phi^{(N)}_1-\phi^{(N)}_2+\phi^{(N)}_3 \quad (\mbox{mod 2} \pi) \label{eq:1/8bps3}  \\
&\mbox{or}& \nonumber \\
& &-\phi^{(1)}_1+\phi^{(1)}_2+\phi^{(1)}_3 = -\phi^{(2)}_1+\phi^{(2)}_2+\phi^{(2)}_3 = \ldots = -\phi^{(N)}_1+\phi^{(N)}_2+\phi^{(N)}_3 (\mbox{mod 2} \pi). \label{eq:1/8bps4} 
\end{eqnarray}

Similarly we obtain 1/4 BPS condition and 1/2 BPS condition. 1/4 BPS condition is
\begin{eqnarray}
& &\phi_1^1=\phi_1^2=\ldots=\phi_1^N, \phi_2^1-\phi_3^1=\phi_2^2-\phi_3^2=\ldots = \phi_2^N-\phi_3^N \quad (\mbox{mod 2} \pi) \label{eq:1/4bps1} \\
&\mbox{or}& \nonumber \\
& &\phi_1^1=\phi_1^2=\ldots=\phi_1^N, \phi_2^1+\phi_3^1=\phi_2^2+\phi_3^2=\ldots = \phi_2^N+\phi_3^N \quad (\mbox{mod 2} \pi). \label{eq:1/4bps2}
\end{eqnarray}
1/2 BPS condition is
\begin{equation}
\phi_1^1=\phi_1^2=\ldots=\phi_1^N,\phi_2^1=\phi_2^2=\ldots=\phi_2^N,\phi_3^1=\phi_3^2=\ldots=\phi_3^N \quad (\mbox{mod 2} \pi). \label{eq:1/2bps}
\end{equation} 

\section{Torons, Condition of Stability, Twisted Boundary Condition}
We will investigate topologically stable solutions of U(N) super Yang-Mills theory on $T^6$, called torons. We are interested only in diagonal elements of U(N) constant field strength, for which an equation of motion is satisfied trivially. A condition of topological stability for a constant flux $\mathcal{F}$ is ~\cite{taylor}
\begin{equation}
\frac{\delta}{\delta \mathcal{F}_{\mu \nu}^a}(\int d^6 \tr(\mathcal{F}_{\mu \nu}\mathcal{F}^{\mu \nu})-\lambda( \frac{1}{3!(2\pi)^3}\int d^6x \tr(\mathcal{F} \wedge \mathcal{F} \wedge \mathcal{F})-n))=0, \label{top}
\end{equation}
where $\lambda$ is a lagrange multiplier, to be determined. $n$ is an instanton number, which is a D0-brane charge in the string point of view. We set a D6-brane tension one and also set a D6-brane world volume one. The action has a local minimum as a functional of $\mathcal{F} =  F + B $ when a B field is turned on. Equation (\ref{top}) leads to 
\[\mathcal{F}_{12}^a = \lambda \frac{1}{2(2\pi)^3}\mathcal{F}_{34}^a\mathcal{F}_{56}^a \]  
\[\mathcal{F}_{34}^a = \lambda \frac{1}{2(2\pi)^3}\mathcal{F}_{56}^a\mathcal{F}_{12}^a \]  
\[\mathcal{F}_{56}^a = \lambda \frac{1}{2(2\pi)^3}\mathcal{F}_{12}^a\mathcal{F}_{34}^a \] 
where we do not sum over $a$.
The stability condition reduces to
\begin{equation}
(\mathcal{F}_{12}^a)^2=(\mathcal{F}_{34}^a)^2=(\mathcal{F}_{56}^a)^2, \label{eq:stability}
\end{equation} 
for each $a$. A boundary condition for a constant gauge field $F$ needs a twist, which is a boundary gauge transformation ~\cite{thooft}. On the other hand, a B field does not need a boundary twist because it is a potential field and a boundary condition of constant B is trivial. We denote short-hand notation for functions defined on a wall of $T^6$:
\begin{equation}
f(x_1=0) \equiv f(x_1=0,x_2,x_3,x_4,x_5,x_6),
\end{equation}
etc. The gauge field obeys a following boundary condition,
\[A_\nu(x_\mu=1)= \Omega_\mu(A_\nu-i\frac{\partial}{\partial x_\nu})\Omega_\mu^{-1}(x_\mu=0). \] 
For the ansatz (\ref{eq:ansatz}), boundary twists of $2i-1$ and $2i$ directions are  
\[\Omega_{2i-1} = \exp(iF_{2i-1 \; 2i}x_{2i}), \quad \Omega_{2i}=1,\]
where $i$ runs from 1 to 3. We do not sum over $i$. These twists need to satisfy the cocycle condition ~\cite{thooft},
\[\Omega_{\mu}(x_{\nu}=1)\Omega_{\nu}(x_{\mu}=0) = \Omega_{\nu}(x_{\mu}=1)\Omega_{\mu}(x_{\nu}=0) Z, \]
where $Z$ is a center of U(N), that is $ \exp(i\theta). $
The above condition reduces to a flux quantization condition:
\begin{equation}
\exp(iF^1_{2i-1 \; 2i}) = \exp(iF^2_{2i-1 \; 2i}) = \ldots = \exp(iF^N_{2i-1 \; 2i}) = \exp(i\theta), \label{eq:fluxquantize}
\end{equation}
for each $i$. These boundary twists suit to solutions which we will construct later. Then, a boundary condition for gauge potential $ A $ is
\begin{eqnarray*}
A_{2i-1}(x_\mu = 1) &=& A_{2i-1}(x_\mu = 0) \\
A_{2i}(x_{2i-1} = 1) &=& A_{2i}(x_{2i-1} = 0) + iF_{2i-1 \; 2i} \\
A_{2i}(x_\mu = 1) &=& A_{2i}(x_\mu = 0), \mbox{ where } \mu \neq 2i-1.  
\end{eqnarray*}
With the above boundary condition, a gauge potential whose strength is constant is
\[A_{2i-1} = 0, \qquad  A_{2i} = i F_{2i-1 \; 2i}x_{2i-1}. \]
Its strength has the form of the ansatz (\ref{eq:ansatz}).

\section{Brane Charge, Energy of Bound States, Energy of Localized states}
Because there are N D6-branes, D6-brane charge is N. The lower dimensional brane charges can be read off by the Chern-Simons-like term of the string effective action,
\[i \mu_p \int_{p+1} \tr(\exp(F_2 + B_2) \wedge \Sigma_q C_q). \]
Then, a D4-charge is 
\[\frac{1}{2\pi}\int_2\tr(\mathcal{F}). \]
In $v^a$ basis, it is
\begin{equation}
 \begin{array}{cc}
(3,4,5,6) \mbox{ D4 charge } & \frac{1}{2\pi} \Sigma_{a=1}^N \mathcal{F}_{12}^a \\
(1,2,5,6) \mbox{ D4 charge } & \frac{1}{2\pi} \Sigma_{a=1}^N \mathcal{F}_{34}^a \\
(1,2,3,4) \mbox{ D4 charge } & \frac{1}{2\pi} \Sigma_{a=1}^N \mathcal{F}_{56}^a.  
\end{array} \label{d4charge}
\end{equation}
These are charges of a U(1) B-field. A D2-brane charge is
\[\frac{1}{2(2\pi)^2}\int_4\tr(\mathcal{F} \wedge \mathcal{F}).\]
Then, it is explicitly,
\begin{equation}
 \begin{array}{cc}
(1,2) \mbox{ D2 charge } & \frac{1}{(2\pi)^2} \Sigma_{a=1}^N \mathcal{F}_{34}^a \mathcal{F}_{56}^a \\
(3,4) \mbox{ D2 charge } & \frac{1}{(2\pi)^2} \Sigma_{a=1}^N \mathcal{F}_{56}^a \mathcal{F}_{12}^a \\
(5,6) \mbox{ D2 charge } & \frac{1}{(2\pi)^2} \Sigma_{a=1}^N \mathcal{F}_{12}^a \mathcal{F}_{34}^a.
\end{array} \label{d2charge}
\end{equation}
A D0-brane charge is 
\[\frac{1}{3!(2\pi)^3}\int_6 \tr(\mathcal{F} \wedge \mathcal{F} \wedge \mathcal{F}),\]
that is
\begin{equation}
\frac{1}{(2\pi)^3}\Sigma_{a=1}^N \mathcal{F}_{12}^a \mathcal{F}_{34}^a \mathcal{F}_{56}^a.  \label{d0charge}
\end{equation}
Now, we can understand physical meaning of the BPS condition and the stability condition (\ref{eq:stability}). We can rewrite the 1/8 BPS condition (\ref{eq:1/8bps1}) in terms of the gauge fields,
\begin{eqnarray}
\frac{\mathcal{F}^1_{12}+\mathcal{F}^1_{34}+\mathcal{F}^1_{56}-\mathcal{F}^1_{12}\mathcal{F}^1_{34}\mathcal{F}^1_{56}}{1-(\mathcal{F}^1_{12}\mathcal{F}^1_{34}+\mathcal{F}^1_{34}\mathcal{F}^1_{56}+\mathcal{F}^1_{56}\mathcal{F}^1_{12})} &=& \frac{\mathcal{F}^2_{12}+\mathcal{F}^2_{34}+\mathcal{F}^2_{56}-\mathcal{F}^2_{12}\mathcal{F}^2_{34}\mathcal{F}^2_{56}}{1-(\mathcal{F}^2_{12}\mathcal{F}^2_{34}+\mathcal{F}^2_{34}\mathcal{F}^2_{56}+\mathcal{F}^2_{56}\mathcal{F}^2_{12})} \nonumber \\
&=& \ldots \nonumber \\
&=& \frac{\mathcal{F}^N_{12}+\mathcal{F}^N_{34}+\mathcal{F}^N_{56}-\mathcal{F}^N_{12}\mathcal{F}^N_{34}\mathcal{F}^N_{56}}{1-(\mathcal{F}^N_{12}\mathcal{F}^N_{34}+\mathcal{F}^N_{34}\mathcal{F}^N_{56}+\mathcal{F}^N_{56}\mathcal{F}^N_{12})}.
\end{eqnarray}
This is the necessary condition for (\ref{eq:1/8bps1}). This relates the lower dimensional brane charges of one D6-brane to those of other D6-branes. 1/4 BPS conditions (\ref{eq:1/4bps1})(\ref{eq:1/4bps2}) are
\begin{eqnarray}
& &\mathcal{F}_{12}^{(1)}=\mathcal{F}_{12}^{(2)}=\ldots=\mathcal{F}_{12}^{(N)}, \quad \frac{\mathcal{F}_{34}^{(1)}-\mathcal{F}_{56}^{(1)}}{1+\mathcal{F}_{34}^{(1)}\mathcal{F}_{56}^{(1)}}=\frac{\mathcal{F}_{34}^{(2)}-\mathcal{F}_{56}^{(2)}}{1+\mathcal{F}_{34}^{(2)}\mathcal{F}_{56}^{(2)}}=\ldots=\frac{\mathcal{F}_{34}^{(N)}-\mathcal{F}_{56}^{(N)}}{1+\mathcal{F}_{34}^{(N)}\mathcal{F}_{56}^{(N)}} \nonumber \\
&\mbox{or}& \nonumber \\
& &\mathcal{F}_{12}^{(1)}=\mathcal{F}_{12}^{(2)}=\ldots=\mathcal{F}_{12}^{(N)}, \quad \frac{\mathcal{F}_{34}^{(1)}+\mathcal{F}_{56}^{(1)}}{1-\mathcal{F}_{34}^{(1)}\mathcal{F}_{56}^{(1)}}=\frac{\mathcal{F}_{34}^{(2)}+\mathcal{F}_{56}^{(2)}}{1-\mathcal{F}_{34}^{(2)}\mathcal{F}_{56}^{(2)}}=\ldots=\frac{\mathcal{F}_{34}^{(N)}+\mathcal{F}_{56}^{(N)}}{1-\mathcal{F}_{34}^{(N)}\mathcal{F}_{56}^{(N)}}. \nonumber \\
\end{eqnarray}
1/2 BPS condition (\ref{eq:1/2bps}) is
\[
\mathcal{F}_{12}^{(1)}=\mathcal{F}_{12}^{(2)}=\ldots=\mathcal{F}_{12}^{(N)}, \quad \mathcal{F}_{34}^{(1)}=\mathcal{F}_{34}^{(2)}=\ldots=\mathcal{F}_{34}^{(N)}, \quad \mathcal{F}_{56}^{(1)}=\mathcal{F}_{56}^{(2)}=\ldots=\mathcal{F}_{56}^{(N)}. 
\] 
 The BPS condition relates the gauge fields on two or more branes. Whereas the stability condition relates the gauge fields of a single brane.
 
To compare this system with localized brane system, let us calculate the DBI energy ~\cite{dbi}. We know that for accurate energy, we have to calculate DBI energy without Yang-Mills energy ~\cite{born, h-t}.
As we are dealing with only the diagonal part of U(N) Yang-Mills theory, the DBI action is essentially abelian, 
\begin{eqnarray}
E &=& \tr\sqrt{det(1+\frac{1}{2\pi}\mathcal{F})} \nonumber \\
  &=& \tr\sqrt{(1+(\mathcal{F}_{12})^2)(1+(\mathcal{F}_{34})^2)(1+(\mathcal{F}_{56})^2)} \nonumber \\ 
  &=& \tr(\Sigma_{a=1}^N (1+(\mathcal{F}_{12}^a)^2 +(\mathcal{F}_{34}^a)^2 +(\mathcal{F}_{56}^a)^2 + (\mathcal{F}_{12}^a)^2(\mathcal{F}_{34}^a)^2 + (\mathcal{F}_{34}^a)^2(\mathcal{F}_{56}^a)^2 + (\mathcal{F}_{56}^a)^2(\mathcal{F}_{12}^a)^2 \nonumber \\
  &+& (\mathcal{F}_{12}^a)^2(\mathcal{F}_{45}^a)^2(\mathcal{F}_{56}^a)^2)v^a))^{1/2} \nonumber \\
  &=& \Sigma_{a=1}^N (1+(\mathcal{F}_{12}^a)^2 +(\mathcal{F}_{34}^a)^2 +(\mathcal{F}_{56}^a)^2 + (\mathcal{F}_{12}^a)^2(\mathcal{F}_{34}^a)^2 + (\mathcal{F}_{34}^a)^2(\mathcal{F}_{56}^a)^2 + (\mathcal{F}_{56}^a)^2(\mathcal{F}_{12}^a)^2 \nonumber\\
  &+& (\mathcal{F}_{12}^a)^2(\mathcal{F}_{45}^a)^2(\mathcal{F}_{56}^a)^2)^{1/2}. \label{eq:dbi}
\end{eqnarray}
We have used the $v^a$ basis and the Taylor expansion.

Next, let us calculate energy of the state in which lower dimensional branes are localized on N coincident D6-branes. The energy of each brane is absolute value of charge times brane tension times world volume. Then, energy of the localized state $E_{local}$ is
\begin{eqnarray}
E_{local} &=& N \nonumber \\
        &+& |\Sigma_{a=1}^N \mathcal{F}_{12}^a|+ |\Sigma_{a=1}^N \mathcal{F}_{34}^a|+ |\Sigma_{a=1}^N \mathcal{F}_{56}^a| \nonumber \\
        &+& |\Sigma_{a=1}^N \mathcal{F}_{12}^a\mathcal{F}_{34}^a| +|\Sigma_{a=1}^N \mathcal{F}_{34}^a\mathcal{F}_{56}^a| + |\Sigma_{a=1}^N \mathcal{F}_{56}^a\mathcal{F}_{12}^a| \nonumber \\
        &+& |\Sigma_{a=1}^N \mathcal{F}_{12}^a\mathcal{F}_{34}^a\mathcal{F}_{56}^a|. \label{eq:local}
\end{eqnarray}

\section{Solutions and Their Physical Meaning}
We construct 1/8 BPS and 7/8 SUSY breaking solutions which are topologically stable and are spanned by diagonal elements of U(N) constant field strengths. We  also set a condition that solutions do not satisfy 1/4 nor 1/2 BPS conditions (\ref{eq:1/4bps1}) (\ref{eq:1/4bps2}) (\ref{eq:1/2bps}). 

Under the condition of topological stability (\ref{eq:stability}) and BPS condition (\ref{eq:1/8bps1})we find \footnote{We also find solutions which consist of $(\mathcal{F}_{12}^{(1)},\mathcal{F}_{34}^{(1)},\mathcal{F}_{56}^{(1)})=(2 \pi/3 M, 2 \pi/3 M, 2 \pi/3 M)$ and $(\mathcal{F}_{12}^{(i)},\mathcal{F}_{34}^{(i)},\mathcal{F}_{56}^{(i)})=(2 \pi M,2 \pi M,-2 \pi M) \mbox{ or } (2 \pi M,-2 \pi M,2 \pi M) \mbox{ or } (-2 \pi M,2 \pi M,2 \pi M) $, where $i$ runs 2 to $N$. This and (\ref{eq:1/8sol}) are a complete system under the condition of topological stability (\ref{eq:stability}) and BPS condition (\ref{eq:1/8bps1}). The solutions have the same property as (\ref{eq:1/8sol}).}   

\begin{eqnarray}
(\mathcal{F}_{12}^{(1)},\mathcal{F}_{34}^{(1)},\mathcal{F}_{56}^{(1)})&=&(2 \pi M,2 \pi M,-2 \pi M) \nonumber \\
(\mathcal{F}_{12}^{(2)},\mathcal{F}_{34}^{(2)},\mathcal{F}_{56}^{(2)})&=&(2 \pi M,-2 \pi M,2 \pi M) \nonumber \\
(\mathcal{F}_{12}^{(3)},\mathcal{F}_{34}^{(3)},\mathcal{F}_{56}^{(3)})&=&(-2 \pi M,2 \pi M,2 \pi M) \nonumber \\
(\mathcal{F}_{12}^{(i)},\mathcal{F}_{34}^{(i)},\mathcal{F}_{56}^{(i)})&=&(2 \pi M,2 \pi M,-2 \pi M) \mbox{ or } (2 \pi M,-2 \pi M,2 \pi M) \nonumber \\ 
&&\mbox{ or } (-2 \pi M,2 \pi M,2 \pi M),  \label{eq:1/8sol}
\end{eqnarray}

in case of (\ref{eq:1/8bps2}), 
\begin{eqnarray}
(\mathcal{F}_{12}^{(1)},\mathcal{F}_{34}^{(1)},\mathcal{F}_{56}^{(1)})&=&(2 \pi M,2 \pi M,2 \pi M) \nonumber \\
(\mathcal{F}_{12}^{(2)},\mathcal{F}_{34}^{(2)},\mathcal{F}_{56}^{(2)})&=&(2 \pi M,-2 \pi M,-2 \pi M) \nonumber \\
(\mathcal{F}_{12}^{(3)},\mathcal{F}_{34}^{(3)},\mathcal{F}_{56}^{(3)})&=&(-2 \pi M,2 \pi M,-2 \pi M) \nonumber \\
(\mathcal{F}_{12}^{(i)},\mathcal{F}_{34}^{(i)},\mathcal{F}_{56}^{(i)})&=&(2 \pi M,2 \pi M,2 \pi M) \mbox{ or } (2 \pi M,-2 \pi M,-2 \pi M) \nonumber \\ 
&&\mbox{ or } (-2 \pi M,2 \pi M,-2 \pi M), \label{solution} 
\end{eqnarray}
where $i$ runs from 4 to N. M is a moduli parameter. It is integral because of the flux quantization condition (\ref{eq:fluxquantize}). It is proportional to N because of the charge quantization of lower dimensional branes. We will explicitly find all the charges later. There is no solution in case of $N < 3$. There is freedom to permute $a$ of $\mathcal{F}^{(a)}$. Solutions of the BPS conditions (\ref{eq:1/8bps3}) (\ref{eq:1/8bps4}) are the permutation of $(\mathcal{F}_{12}, \mathcal{F}_{34}, \mathcal{F}_{56} )$ of $(\ref{solution})$. From now on, we will investigate only (\ref{eq:1/8sol}) case for simplicity. U(1) parts of these solutions proportional to $diag(1,1,\ldots,1)$ are
\begin{eqnarray}
B_{12}=\frac{1}{N}2 \pi M(1+\Sigma_{i=4}^N\sigma_{12}^i) \nonumber \\
B_{34}=\frac{1}{N}2 \pi M(1+\Sigma_{i=4}^N\sigma_{34}^i) \nonumber \\
B_{56}=\frac{1}{N}2 \pi M(1+\Sigma_{i=4}^N\sigma_{56}^i), \nonumber 
\end{eqnarray}
where $\sigma_{12}^i$ is a sign of $\mathcal{F}_{12}^{(i)}$, which is a coefficient of M and so on. Charges of lower dimensional branes are

\begin{tabular}{lll}
          & charges of the BPS solution & 'charges from the B field' \\
D4 (3456) & $M(1+\Sigma_{i=4}^N\sigma_{12}^i)$ & $M(1+\Sigma_{i=4}^N\sigma_{12}^i)$ \\
D4 (1256) & $M(1+\Sigma_{i=4}^N\sigma_{34}^i)$ & $M(1+\Sigma_{i=4}^N\sigma_{34}^i)$ \\
D4 (1234) & $M(1+\Sigma_{i=4}^N\sigma_{56}^i)$ & $M(1+\Sigma_{i=4}^N\sigma_{56}^i)$ \\
D2 (12)   & $M^2(-1+\Sigma_{i=4}^N\sigma_{34}^i\sigma_{56}^i)$ & $\frac{1}{N}M^2(1+\Sigma_{i=4}^N\sigma_{34}^i)(1+\Sigma_{i=4}^N\sigma_{56}^i)$ \\
D2 (34)   & $M^2(-1+\Sigma_{i=4}^N\sigma_{12}^i\sigma_{56}^i)$ & $\frac{1}{N}M^2(1+\Sigma_{i=4}^N\sigma_{12}^i)(1+\Sigma_{i=4}^N\sigma_{56}^i)$ \\
D2 (56)   & $M^2(-1+\Sigma_{i=4}^N\sigma_{12}^i\sigma_{34}^i)$ & $\frac{1}{N}M^2(1+\Sigma_{i=4}^N\sigma_{12}^i)(1+\Sigma_{i=4}^N\sigma_{34}^i)$ \\ 
D0        & $M^3(-3+\Sigma_{i=4}^N\sigma_{12}^i\sigma_{34}^i\sigma_{56}^i)$ & $\frac{1}{N^2}M^3(1+\Sigma_{i=4}^N\sigma_{12}^i)(1+\Sigma_{i=4}^N\sigma_{34}^i)(1+\Sigma_{i=4}^N\sigma_{56}^i)$ 
\end{tabular}
The 'charges from the B field' are charges from a B field substituted for $\mathcal{F}$ in (\ref{d4charge}) (\ref{d2charge}) (\ref{d0charge}). What it means will be explained later. 

We construct 1/4 BPS and 3/4 SUSY breaking solutions which are topologically stable and are spanned by diagonal elements of U(N) constant field strengths. We also set the condition that solutions do not satisfy the 1/2 BPS conditions (\ref{eq:1/2bps}). Under the condition of topological stability (\ref{eq:stability}) and BPS condition (\ref{eq:1/4bps1}) (\ref{eq:1/4bps2}) we find 

in case of (\ref{eq:1/4bps1}), 
\begin{eqnarray}
&& \mathcal{F}_{12}^{(1)}=\mathcal{F}_{12}^{(2)}= \ldots = \mathcal{F}_{12}^{(N)}=2 \pi M \nonumber \\
&&(\mathcal{F}_{34}^{(1)},\mathcal{F}_{56}^{(1)})=(\mathcal{F}_{34}^{(2)},\mathcal{F}_{56}^{(2)})=\ldots=(\mathcal{F}_{34}^{(k)},\mathcal{F}_{56}^{(k)})=(2 \pi M,-2 \pi M) \nonumber \\
&&(\mathcal{F}_{34}^{(k+1)},\mathcal{F}_{56}^{(k+1)})=(\mathcal{F}_{34}^{(k+2)},\mathcal{F}_{56}^{(k+2)})=\ldots=(\mathcal{F}_{34}^{(N)},\mathcal{F}_{56}^{(N)})=(-2 \pi M,2 \pi M),  \label{eq:1/4sol1}
\end{eqnarray}
where $k$ runs from 1 to N$-$1, in case of (\ref{eq:1/4bps2}), 
\begin{eqnarray}
&& \mathcal{F}_{12}^{(1)}=\mathcal{F}_{12}^{(2)}= \ldots = \mathcal{F}_{12}^{(N)}=2 \pi M \nonumber \\
&&(\mathcal{F}_{34}^{(1)},\mathcal{F}_{56}^{(1)})=(\mathcal{F}_{34}^{(2)},\mathcal{F}_{56}^{(2)})=\ldots=(\mathcal{F}_{34}^{(k)},\mathcal{F}_{56}^{(k)})=(2 \pi M,2 \pi M) \nonumber \\
&&(\mathcal{F}_{34}^{(k+1)},\mathcal{F}_{56}^{(k+1)})=(\mathcal{F}_{34}^{(k+2)},\mathcal{F}_{56}^{(k+2)})=\ldots=(\mathcal{F}_{34}^{(N)},\mathcal{F}_{56}^{(N)})=(-2 \pi M,-2 \pi M),  \label{eq:1/4sol2}
\end{eqnarray}
where $k$ runs from 1 to N$-$1. M is the moduli parameter which is integral and proportional to N. There is no solution in case of $N < 2$. There are freedom to change the permutation of $a$ of $\mathcal{F}_{\mu\nu}^{(a)}$. There are also freedom to change the permutation of $(\mathcal{F}_{12}, \mathcal{F}_{34}, \mathcal{F}_{56} )$. From now on, we will investigate only (\ref{eq:1/4sol1}) case for simplicity. The U(1) parts of these solutions proportional to $diag(1,1,\ldots,1)$ are
\begin{eqnarray}
&&B_{12}=2 \pi M \nonumber \\
&&B_{34}=\frac{2k-N}{N}2 \pi M \nonumber \\
&&B_{56}=\frac{N-2k}{N}2 \pi M. \nonumber 
\end{eqnarray}
Charges of lower dimensional branes are

\begin{tabular}{lll}
          & charges of the BPS solution & 'charges from the B field' \\
D4 (3456) & $NM$ & $NM$ \\
D4 (1256) & $(2k-N)M$ & $(2k-N)M$ \\
D4 (1234) & $(N-2k)M$ & $(N-2k)M$ \\
D2 (12)   & $-NM^2$ & $\frac{-(N-2k)^2}{N}M^2$ \\
D2 (34)   & $(N-2k)M^2$ & $(N-2k)M^2$ \\
D2 (56)   & $(2k-N)M^2$ & $(2k-N)M^2$ \\
D0        & $-NM^3$ & $\frac{-(N-2k)^2}{N}M^3$. \\
\end{tabular}

1/2 BPS solutions which are topologically stable and are spanned by diagonal elements of U(N) constant field strengths are
\begin{eqnarray}
&& \mathcal{F}_{12}^{(1)}=\mathcal{F}_{12}^{(2)}= \ldots = \mathcal{F}_{12}^{(N)}=2 \pi M \nonumber \\
&& \mathcal{F}_{12}^{(1)}=\mathcal{F}_{12}^{(2)}= \ldots = \mathcal{F}_{12}^{(N)}= \pm 2 \pi M \nonumber \\
&& \mathcal{F}_{12}^{(1)}=\mathcal{F}_{12}^{(2)}= \ldots = \mathcal{F}_{12}^{(N)}= \pm 2 \pi M. \label{eq:1/2sol}
\end{eqnarray}
U(1) parts of these solutions proportional to $diag(1,1,\ldots,1)$ are
\begin{eqnarray*}
B_{12} &=& 2 \pi M =\mathcal{F}_{12}^a \\
B_{34} &=& \pm 2 \pi M = \mathcal{F}_{34}^a \\
B_{56} &=& \pm 2 \pi M = \mathcal{F}_{56}^a. 
\end{eqnarray*}
These solutions imply that the 1/2 BPS states are simply N coincident D6-branes in a B field. 

Where do these bound states come from? Let us consider a state decaying to the BPS solutions in this paper. The BPS final state is composed of N coincident D6-branes in which lower dimensional branes have dissolved. On the other hand, the initial state is made of unstable localized lower dimensional D-branes on N coincident D6-branes in a B field. The mechanism for the decay is not clear, but D-brane charges must be conserved in this process. For simplicity, consider an initial state without localized D4-branes. In D6-branes $B_{12}$ equals $2\pi /N$ times (3456) D4-brane charge. $B_{34}$ and $B_{56}$ satisfy the same relation. Then, the B field in the initial state equals that in the final BPS state. The 'charges from the B field' are equivalent to the lower dimensional brane charges induced on D6-branes in the initial states by turning on a B field. Can localized D0-D6-brane systems with a B field decay to these states? The answer is No. The reason is as follows. We suppose that this process can occur. Localized D0-D6-brane systems with a B field is unstable when it is strong. This state would decays to a stable BPS one. D2-brane and D4-brane charges of localized D0-D6-brane systems in a B field are induced solely by the B field. D2-brane charge conservation imply that D2-brane charges of the BPS state and D2-brane 'charges coming from the B field' must coincide. But those charges do not coincide in any BPS states. We prove this fact in the appendix. The interpretation consistent with charge conservation is that D0-D2-D6-brane systems in a B field decay to these 1/8 or 1/4 BPS states. The initial D0-D2-D6 system in a B field is unstable. There are three kind of strings which connect two branes; D0-D2, D0-D6, D2-D6 strings. If we turn on a B field,  D0-D2 string and D2-D6 string are always tachyonic, while D0-D6 strings may or may not be tachyonic, depending on the B field. The D0-D2 string makes the D0-brane dissolve in the D2-brane and the D2-D6 string makes this dissolve in the D6-brane. Consequently, final states are the BPS solutions.            

Now, let us calculate energy of this BPS system, and compare it with the energy of the state in which D0, D2, and D4-branes are localized on D6-branes. From now on, we use $ 2\pi \sqrt{\alpha'} = 1 $ convention. That is, a ratio of  tensions of branes of different dimensions equals one. The energy of the U(N) BPS system is
\begin{equation}
E_{BPS}=N\sqrt{1+3M^2+3M^4+M^6}.
\end{equation} 
This is universal. The coefficient 3 in front of $M^2$ and $M^4$ originates from three directions in (1,2), (3,4) and (5,6). This is independent of the number of unbroken super symmetry generators, and is depend only on moduli parameter M. Next, let us calculate energy of the localized system. For simplicity, we will consider the 1/8 BPS 7/8 SUSY breaking U(3) solution. 
\begin{equation}
E_{local}=3+3|M|+3M^2+3|M|^3. 
\end{equation}
Then,
\begin{equation}
E_{local}^2-E_{BPS}^2(N=3) = 18|M|^5+36|M|^3+18|M|>0,
\end{equation}
which implies that these solutions are bound states. What about the possibility that the initial state is D0-D2-D6-brane system in a B field ? Energy of this system is 
\begin{equation}
E_{D0-D2-D6}= 3 \sqrt{det_6 (1+B)} + \Sigma_{i=1}^{3}|q_2^i| \sqrt{det_2(1+B^i)} +|q_0|. \label{eq:0-2-6}
\end{equation}
The $i$ denotes the $(2i-1,2i)$ direction. $q_2^i$ and $q_0$ are D2-brane charges of $(2i-1,2i)$ directions and D0-brane charges respectively. $det_6$ and $det_2$ are a determinant of a D6-brane world volume and one of a D2-brane world volume. The D6-branes and the D2-branes are in a B field. From the charge conservation law we obtain
\begin{eqnarray*}
&\mbox{ D2 : }& -M^2  = q_2^i + 1/3M^2 \\
&\mbox{ D0 : }& -3M^3 = q_0 + 1/9M^3 + 3M.
\end{eqnarray*}
The left hand side is the charges of the final state. Hence the right hand side is charges of the initial state, and they consist of localized D-brane charges and induced brane charges in the higher dimensional D-branes due to the non-vanishing B field. From the above relation, $q_2^i=-4/3M^2$ and $q_0=-28/9M^3-3M.$ From this (\ref{eq:0-2-6}) becomes
\begin{equation}
E_{D0-D2-D6}= (3+13/3M^2)\sqrt{1+1/9M^2} + (3+28/9M^2)|M|.
\end{equation}
Comparing the energy of the initial state with that of the final state, we find,
\[
E_{D0-D2-D6}^2-E_{BPS}^2 = 224/81M^6+40/3M^2+|M|\sqrt{1+1/9M^2}(728/27M^4+134/3M^2+18) > 0.
\]
This implies that the initial state decay to the final one.

 The solutions which we have obtained are stable. The stability is guaranteed by topology, super symmetry, and energy.

\section{Conclusion}
We have constructed BPS solutions spanned by diagonal elements of U(N) constant field strengths in 6+1 dimensional U(N) super Yang-Mills theory on $T^6$. These solutions correspond to stable D0-D2-D4-D6-brane bound states. Stability is guaranteed by energy, super symmetry, and topology. Unstable D0-D2-D6-brane systems in which D0-branes and D2-branes are localized in a B field would decay to these BPS solutions. This scenario is consistent with brane charge conservation. These solutions should serve as a clue to understand the fate of unstable brane systems. Localized D0-D6-brane systems in a B field can not decay to these simple solutions. The meaning of this fact is not clear. If we drop the condition of topological stability and only impose the BPS condition on a gauge fields, we might be able to construct solutions to which D0-D6-brane systems in a B field decay. It is interesting to examine a solution spanned by off-diagonal parts of U(N) constant fluxes as well. 

\vspace{2cm}

\textbf{\large Acknowledgments}

\vspace{5mm}
We would like to thank specially H. Itoyama not only for useful discussions but also for encouraging me. We also would like to thank Y. Hosotani, T. Matsuo, K. Murakami, N. Ohta, T. Suyama, and A. Tsuchiya for useful discussions.

\newpage

\vspace{2cm}

\textbf{\large Appendix}

\vspace{5mm}

In this appendix, we will prove that D0-D6-brane systems in a B field can not decay to the BPS D0-D2-D4-D6 bound states found in this paper. As we commented in section 5, we only need to prove that D2-brane charge of the BPS solution and 'D2-brane charge from the B field' do not coincide with each other. We will search solutions whose D2-brane charges of $\mathcal{F}$ and of B coincide. One can see that the 1/4 BPS solutions do not satisfy the above condition apparently,
\begin{eqnarray}
&& \mbox{(12) D2-brane charge: } \qquad   -NM^2 \neq \frac{-(N-2k)^2}{N}M^2   \mbox{ ( under the condition (\ref{eq:1/4bps1}) )}, \nonumber \\
&& \mbox{(12) D2-brane charge: } \qquad  NM^2 \neq \frac{(-N+2k)^2}{N}M^2   \mbox{ ( under the condition (\ref{eq:1/4bps2}) )}. \nonumber
\end{eqnarray}
Let us go on to the 1/8 BPS solutions. First we analyze the case under the BPS condition (\ref{eq:1/8bps1}). One of the conditions that the D2-brane charges of $\mathcal{F}$ and  B coincide is
\[M^2(-1+\Sigma_{i=4}^N\sigma_{12}^i\sigma_{34}^i) = \frac{1}{N}M^2(1+\Sigma_{i=4}^N\sigma_{12}^i)(1+\Sigma_{i=4}^N\sigma_{34}^i) \quad  \mbox{( (56) D2-brane )} , \]
 where  
\[ (\sigma_{12}^i, \sigma_{34}^i) = (1,1), (1,-1), (-1,1). \]
We denote the number of $(1,-1)$ by $a$, the number of $(-1,1)$ by $b$. Then the number of $(1,1)$ is $N-3-a-b$. The above condition casts in a simple form,
\[ (a+1)(b+1)=0, \]
which can not be satisfied. Under the condition (\ref{eq:1/8bps1}) the charges of $\mathcal{F}$ and B are different. Under the condition (\ref{eq:1/8bps2}) a (5 6) D2-brane charge is the same as under (\ref{eq:1/8bps1}), and the condition for the charge conservation is,
\[M^2(-1+\Sigma_{i=4}^N\sigma_{12}^i\sigma_{34}^i) = \frac{1}{N}M^2(1+\Sigma_{i=4}^N\sigma_{12}^i)(1+\Sigma_{i=4}^N\sigma_{34}^i) \quad \mbox{( (56) D2-brane )} ,\]
 where  
\[ (\sigma_{12}^i, \sigma_{34}^i) = (1,1), (1,-1), (-1,1). \]
This is just the same as in the case (\ref{eq:1/8bps1}). The proof is completed.

\end{document}